\documentclass[prd,floatfix,onecolumn,amsmath,amssymb,floatfix]{revtex4}
\usepackage{graphicx,color,dcolumn,booktabs,bm}
\usepackage{subcaption}
\usepackage{amssymb}
\usepackage{indentfirst}
\usepackage{epsfig}
\usepackage{feynmf}   
\usepackage{epstopdf}   
\usepackage{slashed}  
\usepackage{cases}
\definecolor{maroon}{RGB}{139,25,150}
\usepackage{pdfpages} 
\usepackage{multirow}
\usepackage{float}
\usepackage{graphicx,color,dcolumn,booktabs,bm}
\usepackage[colorlinks, citecolor=blue,anchorcolor=red,menucolor=red, linkcolor=red,filecolor=red,runcolor=red,urlcolor=blue,frenchlinks=red, urlcolor=blue]{hyperref}
\usepackage{orcidlink}

\begin{document}
\preprint{}
\preprint{}
\title{\color{maroon}{Two-body nonleptonic decays of  $\Omega_{b}\rightarrow \Omega_{c}$  beyond tree level}}

\author{Z. Neishabouri$^{a}$\,\orcidlink{0009-0009-0892-384X}}
	\author{K.~Azizi$^{a,b}$\orcidlink{0000-0003-3741-2167}} 
	\email{kazem.azizi@ut.ac.ir} \thanks{Corresponding author} 
	\author{H.~R.~Moshfegh$^{a,c}$\orcidlink{0000-0002-9657-7116}}
\affiliation{
		$^{a}$Department of Physics, University of Tehran, North Karegar Avenue, Tehran 14395-547, Iran\\
		$^{b}$Department of Physics,  Dogus University, Dudullu-\"{U}mraniye, 34775 Istanbul,  T\"{u}rkiye\\
		$^{c}$Centro Brasileiro de Pesquisas Físicas (CBPF), Rua Dr. Xavier Sigaud,150, URCA, Rio de Janeiro CEP 22290-180, RJ, Brazil}
\date{\today}
\begin{abstract}
We study the nonleptonic decays of $\Omega_{b}\rightarrow
\Omega_{c} P (V)$ with eight pseudoscalar and vector mesons using the naive factorization approach. We analyze all relevant topologies (the tree-level, color-suppressed, and penguin) of these decays and calculate the decay amplitude for each separately.  We determine the decay rates, branching ratios and compare our results with those from other theoretical predictions.  The results obtained may be useful for the analysis of the related data in both ongoing and future experiments.
\end{abstract}
\maketitle
\section{Introduction}
The study of weak decays in hadrons serves as a powerful tool for probing their internal dynamics, testing the interplay between weak and strong interactions, validating models such as the quark model, and extracting standard model (SM) parameters including the Cabibbo-Kobayashi-Maskawa (CKM) matrix elements.
 Heavy hadrons incorporating bottom (b) or charm (c) quarks possess a wide variety of accessible decay channels due to their large masses, and their decays provide particularly valuable information for understanding the underlying interactions in particle physics.
 
The observation of lepton flavor universality (LFU) violation in specific decay channels of B mesons \cite{BaBar:2012obs,LHCb:2014vgu,LHCb:2017vlu} and the discovery of charge conjugation-parity (CP) violation in B meson systems \cite{Belle:2001zzw,Christenson:1964fg,Belle:2001qdd,BaBar:2001ags,Belle:2004nch,BaBar:2004gyj} have significantly increased expectations for detecting new physics beyond the SM in hadrons containing bottom quarks. These findings have driven extensive experimental and theoretical studies focused on heavy hadron decays to search for possible new physics effects that could explain such anomalies.
In recent years, a wide range of decays of heavy hadrons have been observed across multiple experiments, particularly by the LHCb collaboration \cite {ParticleDataGroup:2024cfk,LHCb:2014ofc,LHCb:2014yin,LHCb:2022piu,ATLAS:2015hik,Heredia-DeLaCruz:2011bti,CDF:2009sbo,CDF:2011ac,LHCb:2015qvk,LHCb:2015une}
, providing rich data for analysis. Concurrently, numerous theoretical frameworks have been developed to study non-leptonic and semi-leptonic decays of these hadrons. These include approaches such as factorization methods, light-cone sum rules, QCD sum rules, covariant constituent quark models, and the MIT bag model, among others  \cite{Xu:1992hj,Mohanta:2000nk,Neishabouri:2025abl,Geng:2020ofy,Faustov:2016pal,Singleton:1990ye,Ivanov:1996fj,Khajouei:2025tqw,Aliev:2009jt,Azizi:2008ty}. Together, these experimental and theoretical advances have significantly deepened our understanding of heavy hadron decay dynamics and help to test the SM rigorously.

Compared with semileptonic decays, which involve both leptons and hadrons,  nonleptonic decays featuring exclusively hadronic final states exhibit far more complex QCD dynamics. This arises from the more intricate strong interaction effects within exclusive hadronic environments, as opposed to hybrid leptonic-hadronic configurations. Non-leptonic decays facilitate profound insights into short-distance and long-range dynamical interactions while serving as key arenas for detecting CP violation. CP violation refers to the asymmetry under the combined operations of charge conjugation (C) and parity (P). 
It manifests primarily through direct and indirect mechanisms in weak decays and probes the phases of the CKM matrix in the SM. Direct CP violation
arises from interference between two amplitudes bearing distinct weak and strong phases, where weak phases reverse under CP conjugation while strong phases remain invariant. Within the SM, weak phases originate from the complex entries of the (CKM) matrix, whereas strong phases emerge from hadronic final-state interactions.
It appears as rate asymmetries $A_{CP}$ in decays such as $B^0 \rightarrow K^+\pi^-$ and $B_s^0 \rightarrow K^-\pi^+$, where observed values are 
$-0.0824 \pm 0.0033 \pm 0.0033$ and $0.236 \pm 0.013 \pm 0.011$ respectively \cite{LHCb:2020byh}.
The CP asymmetry for a decay $P\rightarrow f$ is defined as:
\begin{equation}
A_{CP} = \frac{\Gamma(P \to f) - \Gamma(\bar{P} \to \bar{f})}{\Gamma(P \to f) + \Gamma(\bar{P} \to \bar{f})}~.
\end{equation}
This asymmetry becomes nonzero only when at least two contributing amplitudes differ simultaneously in their strong phases  (from final-state rescattering) and weak phases (from CKM factors in the SM).
In nonleptonic B and D decays, interference between tree-level and penguin amplitudes provides the required phase differences. Experiments such as LHCb and Belle determine $A_{CP}$ from ratios of decay rates, finding percent-level asymmetries in channels like $ D^0 \to \pi^+ \pi^-$ \cite{LHCb:2019hro}, which place stringent tests on SM predictions. In baryon decays, for instance $\Lambda_b \to p\pi^-$ \cite{LHCb:2024iis}, asymmetries as large as $O(10\%)$ have been reported, offering a sensitive probe of possible contributions beyond simple tree-dominated dynamics.
The observations mark the first instance of CP violation and a non-zero $A_{CP}$ in baryons, observed in the $\Lambda^0_b \to p K^- \pi^+ \pi^- $decay mode, by $5.2$ standard deviations \cite{LHCb:2025ray}.
Indirect CP violation occurs through mixing in neutral meson systems, such as, $K^0-\bar{K}^0$, where the CP eigenstates differ from the mass eigenstates. It is characterized by the parameter $\epsilon \approx 2.3\times 10^{-3}$\cite{ParticleDataGroup:2024cfk} in the kaon sector, reflecting CP-violating differences in the lifetimes and phases of CP-even and CP-odd components, even in the absence of strong-phase differences in the decay amplitudes.
CP violation constitutes one of the three Sakharov conditions necessary to account for the observed matter-antimatter asymmetry in the universe  \cite{Sakharov:1967dj}. However, its predicted magnitude within the SM remains negligible, implying the potential involvement of additional mechanisms.
Non-leptonic decays emerge as a prime laboratory for probing CP violation through their rich interplay of weak and strong dynamics.

This study focuses on the two-body non-leptonic decay of $\Omega_b$, the heaviest singly heavy baryon, into an $\Omega_c$. While the decay channel $\Omega_b \rightarrow \Omega_c \pi$ has been experimentally observed  with a significance of  $ 3.3 \sigma $ \cite{CDF:2014mon}, other decay modes involving different mesons remain yet to be detected. This, highlights the ongoing exploration and need for further experimental efforts in uncovering additional $\Omega_b$ decay processes.
This decay mode has also been analyzed within various theoretical frameworks.
Two main approaches realize approximate SU(3) flavor symmetry in QCD for two-body non-leptonic decays: The reducible method and the topological method.
Topological diagrams, in particular, provide an intuitive representation of the strong-interaction dynamics embedded in weak decay processes, incorporating both perturbative and nonperturbative QCD effects.
 In two-body non-leptonic baryon decays governed by quark color-flavor dynamics, five distinct topological diagrams contribute via W-boson exchange, classified into reducible tree-level diagrams and irreducible W-exchange diagrams.
 These topologies reflect different mechanisms by which the weak interaction mediates the baryon decay processes. In tree-level diagrams, necessary condition  is that two quarks are shared by the parent and daughter baryons and the W boson is emitted either externally or internally. The difference between them is that the quark produced by the weak transition goes into the final state baryon or meson. These diagrams can be factorized. However, the three different types of W-exchange diagrams are not factorable.
Tree-level processes typically dominate, yet W-exchange contributions can amplify branching ratios by up to $30\%$ in decays like $ \Lambda_b \to \Lambda_c$ \cite{Zhang:2022iun,Ivanov:1997hi}. 

\begin{figure}[H]
\centering
\begin{subfigure}{0.33\textwidth}
    \includegraphics[width=\linewidth]{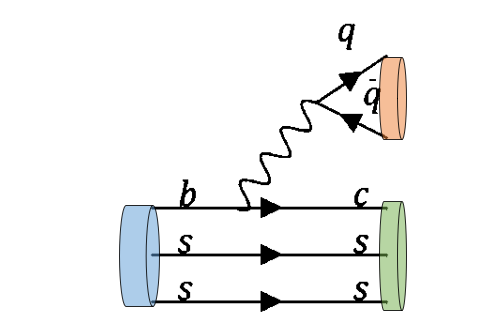}
    \caption*{external W-emision (T)}
  \end{subfigure}
  \begin{subfigure}{0.32\textwidth}
    \includegraphics[width=\linewidth]{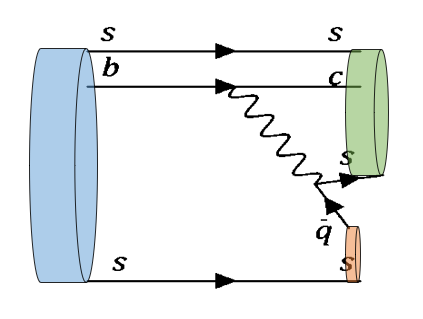}
    \caption*{internal W-emision (C)}
  \end{subfigure}
  \begin{subfigure}{0.33\textwidth}
    \includegraphics[width=\linewidth]{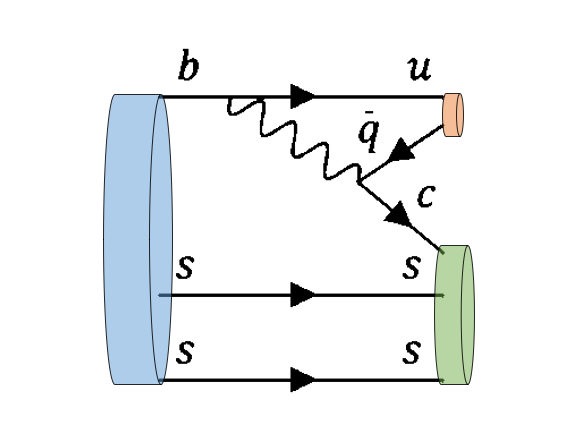}
    \caption*{internal W-emision ($C'$)}
  \end{subfigure}
  \begin{subfigure}{0.32\textwidth}
    \includegraphics[width=\linewidth]{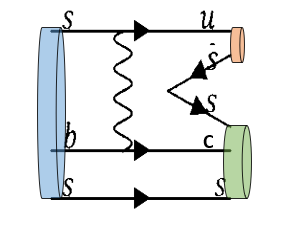}
    \caption*{W-exchange (E)}
  \end{subfigure}
  \begin{subfigure}{0.32\textwidth}
    \includegraphics[width=\linewidth]{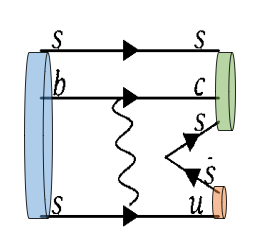}
    \caption*{W-exchange ($E'$)}
  \end{subfigure}
 \caption{All possible flavor-color topologies for  $\Omega_{b}\rightarrow \Omega_{c} M$ decay }\label{topology}
\end{figure}

The simplest model for the non-leptonic decay $\Omega_b \to \Omega_c$ features a $b \to c$ quark transition accompanied by W-boson emission, which subsequently decays into a quark-antiquark pair to form a meson. Although this tree-level emission mode represents the dominant pathway, other scenarios, such as internal W-boson exchange, warrant further investigation to fully characterize the decay dynamics.
 In Fig. \ref{topology}, diagram T represents a color-allowed tree-level process with external W-boson emission, whereas diagrams C and $C^{\prime}$ denote color-suppressed configurations involving internal W emission.
Charge conservation necessitates a negatively charged meson in the final state. Consequently, diagram C permits a strange quark (s) transition, generating kaon $K^{(*)}$, $D_s$ or $D_s^*$ mesons, while diagram $C^{\prime} $remains forbidden due to its inability to produce such a meson.
The $E$ and $E'$ diagrams, involving W-exchange processes, are prohibited by charge conservation. The third W-exchange “bow-tie” topology requires spectator light quark involvement in meson formation, rendering it inaccessible for this decay. Penguin diagrams manifest as higher-order loop contributions within the SM.
In particle physics, penguin topologies refer to specific loop diagrams in weak decays where flavor-changing neutral currents arise through higher-order processes. These loops typically involve virtual particles that mediate transitions between quarks, contributing to rare decay modes and CP violation. Loop diagrams incorporate different internal quarks (e.g., top or charm), giving rise to interference pathways with unique phases. This unitarity driven interference produces differences between decay amplitudes and their CP-conjugate counterparts, rendering CP violation experimentally observable.

In Fig.\ref{penguin}, the penguin diagrams relevant to these decays are shown; they can hadronize into $ D^{(*)}$ and $ D_s^{(*)}$ mesons.  Consequently, their occurrence probability is lower than that of tree-level topologies, and they do not significantly modify the total decay rate.  Nevertheless, they can generate additional complex phases that participate in amplitude interference and therefore play a crucial role in the size of CP violation.
Theoretical analyses of non-leptonic decays rely on factorization theorems, which constitute fundamental tools within perturbative QCD frameworks. These theorems separate physics across different energy scales, distinguishing perturbatively computable short-distance interactions from non-perturbative long-distance effects.
The factorization theorem extends the Euclidean operator product expansion to time-like kinematics, with rigorous proofs derived from perturbative QCD through Feynman diagram analysis in the Collins-Soper-Sterman formalism  \cite{Catani:1989ne,Collins:1981uk,Collins:1984kg}. Heavy quark masses further suppress strong interaction effects through systematic expansions in heavy quark effective theory. Naive factorization was initially applied to B-meson decays and later generalized to heavy-baryon systems. Naive factorization assumes that the hadronic matrix element of a four-quark operator can be written as the product of two single-hadron matrix elements, neglecting nonfactorizable effects such as hard-gluon exchange and spectator interactions.
In this study, we employ the naive factorization approach, which enables a comprehensive calculation of all topologies relevant to these decays.
Form factors have been calculated using the QCD Sum Rules (QCDSR) method, which is one of the powerful and predictive models in the non-perturbative regime of QCD. This method has provided good predictions and results comparable to experiments so far, making it a reliable non-perturbative approach for studying hadronic decays.

This paper is organized as follows. In Sec. \ref{Sec2}, decay amplitudes are derived based on the relevant topological diagrams. In Sec.\ref{Sec3}, decay rates are calculated. Section \ref{Sec4} presents numerical analyses and results, and Sec.  \ref{Sec5} contains the conclusions. Two appendixes  are provided to present some details of the calculations.
\section{The Decay Amplitude}~\label{Sec2}
\begin{figure}
\centering
\begin{subfigure}{0.31\textwidth}
    \includegraphics[width=\linewidth]{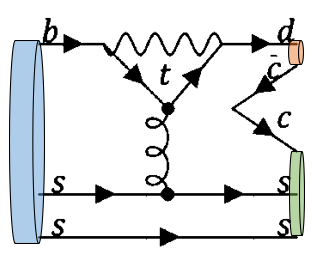}
    \caption*{ W-loop (P)}
  \end{subfigure}
  \begin{subfigure}{0.32\textwidth}
    \includegraphics[width=\linewidth]{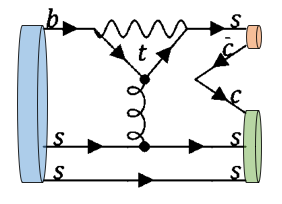}
    \caption*{W-loop(P')}
  \end{subfigure}
 \caption{All possible W-loop diagrams (penguin) for  $\Omega_{b}\rightarrow \Omega_{c} M$ decay }\label{penguin}
\end{figure}

A large category of weak decays in the SM proceed via W-boson exchange; given $m_W \gg m_b\gg m_{\Lambda_{\rm QCD}}$, an effective Hamiltonian describes these processes by integrating out the heavy W boson \cite{Buchalla:1995vs}.
The effective Hamiltonian allows expressing the effects of W boson exchange in terms of local operators and Wilson coefficients. This formalism simplifies the description of weak decays by separating short-distance physics (encoded in Wilson coefficients) from long-distance hadronic effects (described by local operators).
Given the mass difference between $\Omega_{b}$ and $\Omega_{c}$, we analyze eight pseudoscalar and vector mesons that could potentially be produced in this decay. The decay $\Omega_{b}\rightarrow
\Omega_{c} \pi (\rho)$ occurs via external W emission at the quark level through the transition $b\to c \bar{u}d$, as shown by diagram T in Fig.\ref{topology}. 
The effective Hamiltonian for this decay is defined as follows:
\begin{eqnarray}
H_{eff} &=&{G_F\over \sqrt{2}}V_{cb} V^*_{ud} \left \{ C_1(\mu)
O_1 ^d+C_2(\mu) O_2^d  \right \},
\end{eqnarray}
where $G_F$ denotes the Fermi coupling constant, $V_{cb}$ and $V_{ud}$ represent CKM matrix elements, and $C_1, C_2$ are the associated Wilson coefficients.
The operators $O_1$ and $O_2$ constitute four-quark local operators, explicitly defined as:
\begin{equation}\begin{array}{llllll}
O_{1}^d&=&(\bar{d_{i}}u_{i})_{V-A}(\bar{c}_{j}b_{j})_{V-A},  &
O_{2}^d&=&(\bar{d_{i}}u_{j})_{V-A}(\bar{c}
_{j}b_{i})_{V-A},
\end{array}\end{equation}
where  $(\bar q_{1}q_{2})_{V\pm A}=\bar
q_{1}\gamma^{\mu}(1\pm\gamma_{5})q_{2}$ and $ i, j $ denote the color indices. 
To obtain the decay amplitude, the effective Hamiltonian is sandwiched between the initial and final hadron states. This procedure involves calculating the matrix element:
\begin{equation}
	\mathcal{A}(\Omega_{b}\rightarrow\Omega_{c}M) = \langle \Omega_{c} M| \mathcal{H}_{eff} | \Omega_{b}\rangle =\frac{G_{F}}{\sqrt{2}}V_{bc}V_{ud}^{*}\sum_{i=1,2}C_{i}(\mu) \langle \Omega_{c}M| O_{i} | \Omega_{b} \rangle  \,,
\end{equation} 
where $M$ denotes the meson (pseudoscalar or vector). In the next step, using the naive factorization method, the hadronic matrix element is separated into two distinct parts:
\begin{equation}
    \begin{aligned}
     \left\langle \Omega_{c}M\left|O_{i}\right|\Omega_{b}\right\rangle \,\to\langle{M}|\mathcal{J}^{\mu}_{ud}|0\rangle\otimes \langle{\Omega_{c}}|\mathcal{J}_{\mu,cb}|{\Omega_{b}}\rangle \,, \,
    \end{aligned}
\end{equation}
where 
$\mathcal{J}^{\mu}_{ud}$ and $\mathcal{J}_{\mu,cb}$ represent the $  V -A$ currents with the quark flavor $u d$ and $c b$  respectively.
The decay amplitude then takes the form:
\begin{eqnarray}
\mathcal{A}(\Omega_{b}\rightarrow\Omega_{c}M) =&&\frac{G_{F}}{\sqrt{2}}V_{bc}V_{ud}^{*} \bigg\{ C_1(\mu) \langle M|  \bar{d_{i}} \gamma^{\mu}(1-\gamma_{5})u_{i}|0 \rangle \langle \Omega_{c} |\bar{c}_{j} \gamma^{\mu}(1-\gamma_{5}) b_{j} | \Omega_{b}\rangle +\notag\\
&& C_2(\mu)  \langle M|  \bar{d_{i}} \gamma^{\mu}(1-\gamma_{5})u_{j}|0 \rangle \langle \Omega_{c} |\bar{c}_{j} \gamma^{\mu}(1-\gamma_{5}) b_{i} | \Omega_{b}\rangle \bigg\}.
\end{eqnarray}
Using the Fierz identity in color space:
\begin{eqnarray}
&&(T_a)_{ij}(T_a)_{kl}=\frac{1}{2}\big( \delta_{il} \delta_{kj}- \frac{1}{N_c}\delta_{ij}\delta_{kl}\big).\notag\\
\end{eqnarray}
After applying the above relation, the two matrix elements in the decay amplitude become identical. Thus, the amplitude of this decay can be written as:
\begin{eqnarray}\label{eq9}
\mathcal{A}=\left\langle \Omega_{c}M\left|\mathcal{H}_{eff}\right|\Omega_{b}\right\rangle =\frac{G_{F}}{\sqrt{2}}V_{bc}V_{ud}^{*}a_{1}(\mu)\left\langle M\left|\bar{d}\gamma^{\mu}(1-\gamma_{5})u\right|0\right\rangle \left\langle \Omega_{c}\left|\bar{c}\gamma_{\mu}(1-\gamma_{5})b\right|\Omega_{b}\right\rangle,
\end{eqnarray}
 where $a_1(\mu)$ is the effective Wilson coefficient defined through $a_1(\mu) = C_1(\mu) + C_2(\mu)/{N_c}$.
 In Eq.\ref{eq9} the first matrix element corresponds the meson decay constant, while the second matrix element is parameterized in terms of baryonic transition form factors. We now apply this formalism to analyze various decay channels. First we examine the decay amplitude of $\Omega_{b}\rightarrow\Omega_{c} K_u (K_u^*) $ with the quark-level transition $b\to c \bar{u}s$. This 
 decay  can occur via both external (T) and (C) internal W emission topologies, as shown in Fig.\ref{topology}.  
 The effective Hamiltonian describing this decay is expressed as follows:
\begin{eqnarray}
H_{eff} &=&{G_F\over \sqrt{2}}  \left \{V_{cb}  V^*_{us}(C_1(\mu)
O_1 ^s+C_2(\mu) O_2^s) \right \},
\end{eqnarray}
where the four-quark operators are:
\begin{equation}\begin{array}{llllll}
O_{1}^s&=&(\bar{s_{i}}u_{i})_{V-A}(\bar{c}_{j}b_{j})_{V-A},  &
O_{2}^s&=&(\bar{s_{i}}u_{j})_{V-A}(\bar{c}
_{j}b_{i})_{V-A}.
\end{array}\end{equation}
The decay amplitude is obtained by sandwiching the effective Hamiltonian between the initial and the final hadronic states. Using factorization hypothesis and the Fierz identity, the amplitude can be expressed distinctly for the two topologies:
1) For the external W-emission (color-allowed), the amplitude factorizes into the product of a meson decay constant and a baryon transition form factor, reflecting the color-favored topology in which the W boson directly produces the meson.
2) For the internal W-emission (color-suppressed), the amplitude involves color rearrangement via the Fierz transformation, leading to suppression due to color mismatch. This contribution factorizes similarly but is multiplied by $1/{N_c}$ color-suppression factor. Accordingly, the amplitudes can be written as:
\begin{eqnarray}
&&\mathcal{A}_1=\left\langle \Omega_{c}K^{(*)}\left|\mathcal{H}_{eff}\right|\Omega_{b}\right\rangle^{T} =\frac{G_{F}}{\sqrt{2}}V_{bc}V_{us}^{*}a_{1}(\mu)\left\langle K^{(*)}\left|\bar{s}\gamma^{\mu}(1-\gamma_{5})u\right|0\right\rangle \left\langle \Omega_{c}\left|\bar{c}\gamma_{\mu}(1-\gamma_{5})b\right|\Omega_{b}\right\rangle, \,\notag\\
&&\mathcal{A}_2=\left\langle \Omega_{c}K^{(*)}\left|\mathcal{H}_{eff}\right|\Omega_{b}\right\rangle^{C} =\frac{G_{F}}{\sqrt{2}}V_{bc}V_{us}^{*}a_{2}(\mu)\left\langle K^{(*)}\left|\bar{s}\gamma^{\mu}(1-\gamma_{5})u\right|0\right\rangle  \left\langle \Omega_{c}\left|\bar{c}\gamma_{\mu}(1-\gamma_{5})b\right|\Omega_{b}\right\rangle, \,
\end{eqnarray}
  where $a_2(\mu)$ is defined as,  $a_2(\mu) = C_2(\mu) + C_1(\mu)/{N_c}$. 
  The next decay is $\Omega_{b}\rightarrow
\Omega_{c} D (D^*) $, which occurs dominantly via external W-emission (T) as shown in Fig.\ref{topology}. In addition to the leading tree-level contribution, this decay channel can receive subdominant contributions from higher-order penguin loop diagrams (P), shown in Fig.\ref{penguin}. These penguin contributions introduce small corrections and possible CP-violating phases but are generally suppressed compared to the tree-level amplitude
The effective Hamiltonian corresponding to this decay is:
  \begin{eqnarray}
H_{eff} &=&{G_F\over \sqrt{2}} \left \{ V_{bc} V^*_{cd}(C_1
O_1^c+C_2 O_2^c) +  V_{tb} V^*_{td} \sum_{n=3}^{10} C_n
O_n\right \},
\end{eqnarray}
where $O_n$ are quark operators and are given by:
\begin{equation}\begin{array}{llllll}
O_{1}^c&=&(\bar{q_{i}^{\prime
}}c_{i})_{V-A}(\bar{c}_{j}b_{j})_{V-A},  &
O_{2}^c&=&(\bar{q_{i}^{\prime }}c_{j})_{V-A}(\bar{c}
_{j}b_{i})_{V-A},   \\\\
O_{3}&=&(\bar{q_{i}^{\prime
}}b_{i})_{V-A}\sum_{q}(\bar{q}_{j}q_{j})_{V-A},  &
O_{4}&=&(\bar{q_{i}^{\prime }}b_{j})_{V-A}\sum_{q}(\bar{q}
_{j}q_{i})_{V-A},  \\\\
O_{5}&=&(\bar{q_{i}^{\prime
}}b_{i})_{V-A}\sum_{q}(\bar{q}_{j}q_{j})_{V+A},  &
O_{6}&=&(\bar{q_{i}^{\prime }}b_{j})_{V-A}\sum_{q}(\bar{q}
_{j}q_{i})_{V+A},   \\\\
O_{7}&=&(\bar{q_{i}^{\prime}}b_{i})_{V-A}\sum_{q}{\frac{3}{2}}e_{q}(\bar{q}
_{j}q_{j})_{V+A},&
O_{8}&=&(\bar{q_{i}^{\prime
}}b_{j})_{V-A}\sum_{q}{\frac{3}{2}}e_{q}(\bar{q}_{j}q_{i})_{V+A},\\\\
O_{9}&=&(\bar{q_{i}^{\prime}}b_{i})_{V-A}\sum_{q}{\frac{3}{2}}e_{q}(\bar{q}
_{j}q_{j})_{V-A}  &
 O_{10}&=&(\bar{q_{i}^{\prime
}}b_{j})_{V-A}\sum_{q}{\frac{3}{2}}e_{q}(\bar{q}_{j}q_{i})_{V-A}.\\\\
\end{array}\end{equation}
Here, the sum over $q$ runs over all quark flavors, and $q^{\prime}$ corresponds to d and s for the final-state mesons $D^{(*)}$ and $D_s^{(*)}$,  respectively.
 The effective Hamiltonian operators are categorized as follows:
 $O_1$
 and $O_2$ are current-current operators representing tree-level four-quark interactions. $O_3-O_6$ are QCD penguin operators accounting for gluonic loop corrections.
 $O_7$ through $O_{10}$  are electroweak penguin operators that include photon and $Z$-boson mediated loop effects.

 The amplitude of  decay to $ D $ meson, as an example,  is expressed as:
 \begin{eqnarray}
&&\mathcal{A}=\left\langle \Omega_{c}D\left|\mathcal{H}_{eff}\right|\Omega_{b}\right\rangle =\frac{G_{F}}{\sqrt{2}}\bigg[V_{bc}V_{cd}^{*}a_{1}-V_{tb}V_{td}\big(a_4+a_{10}+R (a_6+a_8)\big)\bigg]\left\langle D\left|\bar{d}\gamma^{\mu}(1-\gamma_{5})c\right|0\right\rangle \left\langle \Omega_{c}\left|\bar{c}\gamma_{\mu}(1-\gamma_{5})b\right|\Omega_{b}\right\rangle. \,\notag\\
\end{eqnarray}
For the vector meson ($D^*$), the contributions from the operators $O_5$ to $O_8$ vanish due to the Dirac and color structure of
these operators. The amplitude becomes:
 \begin{eqnarray}
&&\mathcal{A}=\left\langle \Omega_{c}D^{*}\left|\mathcal{H}_{eff}\right|\Omega_{b}\right\rangle =\frac{G_{F}}{\sqrt{2}}\bigg[V_{bc}V_{cd}^{*}a_{1}-V_{tb}V_{td}\big(a_4+a_{10})\big)\bigg]\left\langle D^{*}\left|\bar{d}\gamma^{\mu}(1-\gamma_{5})c\right|0\right\rangle \left\langle \Omega_{c}\left|\bar{c}\gamma_{\mu}(1-\gamma_{5})b\right|\Omega_{b}\right\rangle, \,\notag\\
\end{eqnarray}
 where Wilson coefficients are obtained as:
\begin{equation}
a_{2i -1}=C_{2i-1}+\frac{1}{N_c}C_{2i},~~~
a_{2i}=C_{2i}+\frac{1}{N_c}C_{2i-1},~~~~
i=1,2 \cdots 5\;,
\end{equation}
with
\begin{equation}
R=\frac{2 m_{D_{q'}}^2}{(m_b-m_c)(m_{q'}+m_c)}\;.
\end{equation}
 The derivation of the penguin-operator coefficients is presented in the Appendix A.
The last decay of $\Omega_{b}\rightarrow
\Omega_{c} D_s (D_s^*) $ with the quark-level transition $b\to c \bar{c}s$ can proceed via both external and internal W-emission diagrams, where the W boson is emitted outside or inside the quark lines, respectively. Additionally, this decay can also occur through penguin loop diagrams ($P^{\prime}$) in Fig.\ref{penguin}, which involve flavor-changing neutral current processes mediated by virtual loops.
The effective Hamiltonian for this decay at the quark level can be written as:
\begin{eqnarray}
H_{eff} &=&{G_F\over \sqrt{2}} \left \{ V_{bc} V^*_{cs}(C_1
O_1^c+C_2 O_2^c ) +  V_{tb} V^*_{ts} \sum_{n=3}^{10} C_n
O_n\right \}.
\end{eqnarray}
The color-allowed (T) and color-suppressed (C) amplitudes in Fig.\ref{penguin}, denoted  by
$\mathcal{A}_1$ and $\mathcal{A}_2$ respectively, can be written as:
 \begin{eqnarray}
&&\mathcal{A}_1=\left\langle \Omega_{c}D_s\left|\mathcal{H}_{eff}\right|\Omega_{b}\right\rangle =\frac{G_{F}}{\sqrt{2}}\big[V_{bc}V_{cs}^{*}a_{1}-V_{tb}V_{ts}\big(a_4+a_{10}+R (a_6+a_8)\big)\big]\left\langle D_s\left|\bar{s}\gamma^{\mu}(1-\gamma_{5})c\right|0\right\rangle \left\langle \Omega_{c}\left|\bar{c}\gamma_{\mu}(1-\gamma_{5})b\right|\Omega_{b}\right\rangle, \,\notag\\
&&\mathcal{A}_1=\left\langle \Omega_{c}D_s^{*}\left|\mathcal{H}_{eff}\right|\Omega_{b}\right\rangle =\frac{G_{F}}{\sqrt{2}}\big[V_{bc}V_{cs}^{*}a_{1}-V_{tb}V_{ts}\big(a_4+a_{10})\big]\left\langle D_s^{*}\left|\bar{s}\gamma^{\mu}(1-\gamma_{5})c\right|0\right\rangle \left\langle \Omega_{c}\left|\bar{c}\gamma_{\mu}(1-\gamma_{5})b\right|\Omega_{b}\right\rangle, \,\notag\\
&&\mathcal{A}_2=\left\langle \Omega_{c}D_s^{(*)}\left|\mathcal{H}_{eff}\right|\Omega_{b}\right\rangle =\frac{G_{F}}{\sqrt{2}}V_{bc}V_{cs}^{*}a_{2}\left\langle D_s^{(*)}\left|\bar{s}\gamma^{\mu}(1-\gamma_{5})c\right|0\right\rangle  \left\langle \Omega_{c}\left|\bar{c}\gamma_{\mu}(1-\gamma_{5})b\right|\Omega_{b}\right\rangle. \,
\end{eqnarray}
The combination of the Wilson coefficients is obtained in the same way as in the previous decays. The decay amplitude in all the above decays is expressed as a product of the Wilson coefficients, the CKM matrix elements, and the Fermi coupling constant, arranged within two matrix elements.
By definition, the matrix element representing the coupling of the meson current to the vacuum is called the decay constant. For pseudoscalar mesons $P(p)$, it is defined as:
\begin{align} \label{pse}
 & \left\langle P(q)\left|\bar{q_i}\gamma^{\mu}(1-\gamma_{5})q'_i\right|0\right\rangle =if_{P}q^{\mu},
 \end{align}
 while for vector mesons $V(q)$, it is given by:
 \begin{align}
 & \left\langle V(q)\left|\bar{q_i}\gamma^{\mu}(1-\gamma_{5})q'_i\right|0\right\rangle =m_{V}f_{V}\epsilon^{\ast\mu} ,
\end{align}
where $f_P$ and $f_V$ are the decay constants of the pseudoscalar and vector mesons, respectively. $m_V$ and $q^\mu$ denote the mass and 
four-momentum of the vector  and pseudoscalar mesons, respectively, $\epsilon^\mu$ is the polarization vector of the vector meson. These parameters quantify the overlap of
the meson wave function at the origin and are crucial inputs for factorization-based amplitude calculations in heavy baryon decays.
The matrix element representing the transition between the initial and the final baryon
states is expressed in terms of transition form factors, which encapsulate the non-perturbative QCD dynamics of the baryon. These form factors can be calculated using various theoretical approaches such as lattice QCD, QCD sum rules or quark models.
The matrix element receives contributions from both vector and axial-vector currents, each parameterized by three independent form factors. The general expression for the baryon transition matrix element is given by:
\begin{eqnarray}\label{Cur.with FormFac.}
&&\langle \Omega_c(p',s')|V^{\mu}|\Omega_b (p,s)\rangle = \bar
u_{\Omega_c}(p',s') \Big[F_1(q^2)\gamma^{\mu}+F_2(q^2)\frac{p^{\mu}}{m_{\Omega_b}}
+F_3(q^2)\frac{p'^{\mu}}{m_{\Omega_c}}\Big] u_{\Omega_b}(p,s), \notag \\
&&\langle \Omega_c(p',s')|A^{\mu}|\Omega_b (p,s)\rangle = \bar u_{\Omega_c}(p',s') \Big[G_1(q^2)\gamma^{\mu}+G_2(q^2)\frac{p^{\mu}}{m_{\Omega_b}}+G_3(q^2)\frac{p'^{\mu}}{m_{\Omega_c}}\Big]
\gamma_5 u_{\Omega_b}(p,s).\notag \\
\end{eqnarray}
In this article, we employ the form factors from our previous work \cite{Neishabouri:2024gbc}. In this study, the form factors have been calculated up to sixth-order non-perturbative contributions (dimension-6 condensates) and the decay form factors for $\Omega_b \to \Omega_c \ell \bar{\nu}_\ell$ were calculated for the first time using QCD sum rules, and were found to be in excellent agreement
with results from other theoretical methods.
The decay amplitude for a pseudoscalar meson can be written using the baryonic transition matrix element (\ref{Cur.with FormFac.}), the pseudoscalar meson matrix element (\ref{pse}), the Dirac equation and  the relation $q=p-p'$ as follows:
\begin{eqnarray}
&&\mathcal{A}_{\Omega_b \to \Omega_cP}=\frac{G_F}{\sqrt2} f_P C_{\rm eff}~ \bar u_{\Omega_c}
\frac{1}{2 m_{\Omega_b} m_{\Omega_c}}\Bigg\{2 F_1 m_{\Omega_b}(m_{\Omega_b} - m_{\Omega_c}) m_{\Omega_c}+ 
  F_2 m_{\Omega_c} (3 m^2_{\Omega_b} + m^2_{\Omega_c} - q^2) -\notag\\ 
 && F_3 m_{\Omega_b} (m^2_{\Omega_b}+ 3 m^2_{\Omega_c} - 
     q^2) + 
     \bigg[2 G_1 m_{\Omega_b} m_{\Omega_c} (m_{\Omega_b}+ m_{\Omega_c}) + 
     G_3 m_{\Omega_b} (m^2_{\Omega_b} + 3 m^2_{\Omega_c} - q^2) +\notag\\ 
     &&G_2 m_{\Omega_c} (-3 m^2_{\Omega_b} - m^2_{\Omega_c} + q^2\bigg] \gamma_5\Bigg\}u_{\Omega_b}.
\end{eqnarray}
The factorized decay amplitude for a vector meson V can be written as:
\begin{eqnarray}
&&\mathcal{A}_{\Omega_b \to \Omega_cV}=\frac{G_F}{\sqrt2} m_V f_V C_{\rm eff}~\bar u_{\Omega_c}\epsilon^{\ast\mu} (F_1 \gamma_\mu+ F_2 \frac{p_\mu}{ m_{\Omega_b}} + F_3 \frac{p'_\mu}{m_{\Omega_c}}- 
  G_1\gamma_\mu \gamma_5 - G_2 \frac{p_\mu \gamma_5}{m_{\Omega_b}} - G_3\frac{p'_\mu~ \gamma_5}{m_{\Omega_c}}) u_{\Omega_b}.
\end{eqnarray}
The  polarization relations $\sum_{\lambda}~\epsilon^{\ast\mu}_{\lambda}(q)\epsilon^\nu_{\lambda}(q)=-g_{\mu\nu}+\frac{q_\mu q_\nu}{m^2_V}$ will be used when squaring the amplitude. The parameter $C_{\rm eff}$ is a factor that incorporates CKM matrix elements and effective Wilson coefficients, which vary for each decay depending on the topology type. In Table \ref{c_eff}, this factor is provided for each decay channel.  
 In the next section, we present the procedure for obtaining the decay
rate using the  transition amplitudes.
\section{ DECAY RATES}~\label{Sec3}
The decay rate for any two-body decay ($A\to B+C$) is generally obtained from the Fermi's Golden Rule  as follows \cite{Thomson:2013zua}:
\begin{eqnarray}
\Gamma_{fi}=\frac{p_c}{32\pi^2 m_A^2}\int |\mathcal{A}_{fi}|^2 d\Omega,
\end{eqnarray}
where the center of mass frame momentum of final state particles is given by:
 $p_c=\frac{1}{2m_A} \lambda^{\frac{1}{2}}(m_A^2,m_B^{2},m_C^{2})$ 
and integral is taken over the phase space.
The decay rate is obtained by inserting the squared decay amplitude and the masses of the hadrons. 
The differential decay rate can also be written in terms of helicity amplitudes. It is important to note that choosing between the direct decay rate formula and the helicity formalism depends on how the hadronic current is expanded in terms of form factors. In the standard helicity expressions \cite{Faessler:2009xn,Gutsche:2017wag}, the baryonic current is parameterized using $\gamma^\mu$, $\sigma_{\mu\nu}q_{\nu}$ and $q^\mu$; with this choice, contracting with the $q^\mu$ coming from the meson matrix element remove the contribution of the form factor $F_2$ and $G_2$ in pseudoscalar decays, and similarly 
once the vector‑meson polarization is included, $F_3$ and $G_3$ drop out of the decay rate for vector mesons.
In contrast, in the present work the baryonic current is expanded in terms of $\gamma^\mu$, $p^\mu$, and $p^{\prime \mu}$. With this parameterization, all three form factors $F_1, F_2$, and $F_3$ (and analogously $ G_1, G_2, G_3$) enter the scalar combinations that build the decay rate, so none of them drop out. For this reason, one must use the direct expression for the differential decay rate of spin 1/2 to spin 1/2 baryon with the factor $\frac{1}{2s+1}$:
\begin{eqnarray}
\Gamma=\frac{\lambda^{1/2}(m_{\Omega_b}^2,m_{\Omega_c}^2,m_{M^2})}{32\pi m^3_{\Omega_b}}|\mathcal{A}|^2 ,
\end{eqnarray}\label{decayformula}
where
$\lambda(m_{\Omega_{b}}^2,m_{\Omega_{c}}^2,m_M^2)=m_{\Omega_{b}}^4+m_{\Omega_{c}}^4+m_M^4-2m_{\Omega_{b}}^2m_{\Omega_{c}}^2-2m_{\Omega_{b}}^2m_M^2-2m_{\Omega_{c}}^2m_M^2$. Appendix B 
contains explicit expressions for $ |\mathcal{A}|^2 $ for both the pseudoscalar and vector meson cases.
\\
   \begin{table}
  \centering
    	\caption{Types of topologies and the corresponding coefficient for each decay.}
  \begin{tabular}{|c|c|c|}
  \hline
  Decay &Topology& $C_{\rm eff}$\\
  \hline
$\Omega_{b}\rightarrow\Omega_{c} \pi (\rho)$  & T&$ [V_{bc} V_{ud}a_1]$\\
$\Omega_{b}\rightarrow\Omega_{c} K (K^*)$  & T,C& $[V_{bc} V_{us}(a_1+a_2)]$\\
$\Omega_{b}\rightarrow\Omega_{c} D$ & T,P& $[V_{bc} V_{cd} a_1-V_{tb}V_{td}(a_4+a_{10}+R(a_6+a_8))]$\\
$\Omega_{b}\rightarrow\Omega_{c} D^*$ &T,P&  $[V_{bc} V_{cd} a_1-V_{tb}V_{td}(a_4+a_{10})]$\\
$\Omega_{b}\rightarrow\Omega_{c} D_s$ &T,C,$P^{\prime}$ &$[V_{bc} V_{cs}(a_1+a_2)-V_{tb}V_{td}(a_4+a_{10}+R(a_6+a_8))]$\\
$\Omega_{b}\rightarrow\Omega_{c} D_s^*$ &T,C,$P^{\prime}$& $[V_{bc} V_{cs} (a_1+a_2)-V_{tb}V_{td}(a_4+a_{10})]$\\
\hline
\end{tabular}
\label{c_eff}
  \end{table}
\section{Numerical Analysis}\label{Sec4}
In this section, we present the input parameters used to calculate the decay rates. The first quantities appearing in the effective Hamiltonian are the Wilson coefficients.
The Wilson coefficients are solutions of the renormalization group equations that relate the short-distance effects from a high-energy scale to the long-distance effects at a lower energy scale and are scale-dependent. These coefficients have been calculated at next-to-leading order in the scale of the b quark mass,  as given beow:\cite{Buchalla:1995vs,Buras:1992zv,Ciuchini:1993vr,Deshpande:1994pc}
\begin{equation}\begin{array}{llllll}
 C_1 & = & 1.117\ ,  &  C_2 & = & - 0.257\ , \\
 C_3 & = & 0.017\ ,  &  C_4 & = & -0.044\ , \\
 C_5 & = & 0.011\ ,  &  C_6 & = & -0.056\ , \\
 C_7 & = & -1\times 10^{-5}\ , &  C_8 & = & 5\times 10^{-4}\ , \\
 C_9 & = & -0.010\ ,  & C_{10} & = & 0.002\ . \\
\end{array}\end{equation}
The next parameters to be included are the decay constants, which are derived from the matrix elements coupling the meson current to the vacuum. These decay constants are reported in Table \ref{decay costant}. Additional input parameters such as the masses of the hadrons and participating quarks, as well as the elements of the CKM matrix used in the numerical calculations, are presented in Table \ref{inputParameter}.
  \begin{table}[h!]
    \centering
    	\caption{Decay constants of the pseudoscalar and vector mesons }
    	\begin{tabular}{cc|cc}
    		\hline
    		Decay constants~($\mathrm{MeV}$)&Value&Decay constants~($\mathrm{MeV}$)&Value\\
    		\hline
    		$f_\pi$ &$130.2~$  \cite{Narison:2012xy,ParticleDataGroup:2024cfk} &   $f_\rho$  & $218.3~$ \cite{Gutsche:2018utw}\\
    		$f_{K}$ &$155.6~$ \cite{Narison:2012xy,ParticleDataGroup:2024cfk}  &	$f_{K^*}$ &$210~$ \cite{Li:2021kfb}\\
    		$f_{D}$ &$203.7~$ \cite{Narison:2012xy} &	$f_{D^*}$& $245\pm20~$\cite{Becirevic:1998ua}\\
    		$f_{D_s}$ &$246.0~$  \cite{Narison:2012xy}&	$f_{D_s^*}$& $272\pm16~$ \cite{Becirevic:1998ua}\\
    		\hline
    		\hline
    	\end{tabular}
    	\label{decay costant}
    \end{table}

  The fitted function for the form factors obtained in \cite{Neishabouri:2024gbc},
  \begin{equation} \label{fitffunction}
{\cal F}(q^2)=\frac{{\cal
F}(0)}{\displaystyle\left(1-a_1\frac{q^2}{m^2_{\Omega_b}}+a_2
\frac{q^4}{m_{\Omega_b}^4}+a_3\frac{q^6}{m_{\Omega_b}^6}+a_4\frac{q^8}{m_{\Omega_b}^8}\right)}.
\end{equation}
  is used to compute the form factors at the momentum transfer squared, $ q^2$, where it is taken as the squared meson mass. The values of the form factors for each decay channel, evaluated at $q^2 = m_{\text{meson}}^2$, are listed in Table \ref{formfactornon}. These values serve as the essential inputs for accurately calculating the decay rates within the factorization framework using the obtained  amplitudes. 

%
\begin{table}[h!]
\caption{Input parameters used in calculations.}\label{inputParameter}
\begin{tabular}{|c|c|}
\hline 
Parameters                                             &  Values  \\
\hline 
$m_{\Omega_{b}}$                                & $(6045.8\pm0.8)~\mathrm{MeV}$\cite{ParticleDataGroup:2024cfk}\\
$m_{\Omega_{c}}$                               &$(2695.3\pm0.4)~\mathrm{MeV}$\cite{ParticleDataGroup:2024cfk}\\
$m_\pi$                                                &$139.57\mathrm{MeV}$\cite{ParticleDataGroup:2024cfk}\\
$m_\rho$                                                 &$770\mathrm{MeV}$\cite{ParticleDataGroup:2024cfk}\\
$m_K$                                                 & $(493.677\pm0.013)~\mathrm{MeV}$\cite{ParticleDataGroup:2024cfk}\\
$m_{K^*}$                                           &  $(891.67\pm0.26)~\mathrm{MeV}$\cite{ParticleDataGroup:2024cfk}\\
$m_D$                                                  & $(189.5\pm0.4)~\mathrm{MeV}$ \cite{ParticleDataGroup:2024cfk}\\
$m_{D^*}$                                          & $(2010.26\pm0.05)\mathrm{MeV}$ \cite{ParticleDataGroup:2024cfk}\\
$m_{D_s}$                                          & $(1968.35\pm0.07 )\mathrm{MeV}$ \cite{ParticleDataGroup:2024cfk}\\
$m_{D_s^*}$                                     &$(2106.6\pm3.4 )\mathrm{MeV}$ \cite{ParticleDataGroup:2024cfk}\\
$m_{b}$                                                &$(4.18^{+0.03}_{-0.02})~\mathrm{GeV}$\cite{ParticleDataGroup:2024cfk}\\
$m_{c}$                                                &$(1.27\pm0.02)~\mathrm{GeV}$\cite{ParticleDataGroup:2024cfk}\\
$G_{F}$                                                &$1.17\times 10^{-5}~\mathrm{GeV^{-2}}$\cite{ParticleDataGroup:2024cfk}\\
$V_{cb}$                                              &$0.040$\cite{ParticleDataGroup:2024cfk}\\
$V_{ud}$                                              &  $0.974$  \cite{ParticleDataGroup:2024cfk} \\
$V_{us}$                                              &   $0.226$  \cite{ParticleDataGroup:2024cfk} \\
$V_{cd}$                                              &  $0.226$  \cite{ParticleDataGroup:2024cfk}  \\
$V_{cs}$                                              &  $0.973$  \cite{ParticleDataGroup:2024cfk} \\
$V_{td}$                                              &  $0.0086$  \cite{ParticleDataGroup:2024cfk} \\
$V_{ts}$                                              &  $0.0415$  \cite{ParticleDataGroup:2024cfk} \\
$\tau_{\Omega_b} $                               & $ 1.64\pm0.16\times 10^{-12} s$  \cite{ParticleDataGroup:2024cfk}\\
\hline
\end{tabular}
\end{table}
  \begin{table}
      	\centering
      	\caption{Transition form factors of the nonleptonic $\Omega_{b}\rightarrow \Omega_{c}M$ weak decays with emitting a pseudoscalar or vector meson at $q^2=m_{Meson}^{2}$.}
      	\begin{ruledtabular}
      		\begin{tabular}{cccccccc}
      	Transition&$F_{1}(q^{2})$&$F_{2}(q^{2})$&$F_{3}(q^{2})$&$G_{1}(q^{2})$&$G_{2}(q^{2})$&$G_{3}(q^{2})$\\
      	\hline
 $\Omega_{b}\rightarrow \Omega_{c}\pi^{-}$&$-0.290\pm0.080$&$0.680\pm0.200$&$-0.038\pm0.012$&$-0.020\pm0.006$&$0.41\pm0.10$&$0.450\pm0.120$\\
 $\Omega_{b}\rightarrow \Omega_{c}K^{-}$&$-0.293\pm0.081$&$0.686\pm0.202$&$-0.039\pm0.012$&$-0.020\pm0.007$&$0.414\pm0.110$&$0.454\pm0.120$\\
      			$\Omega_{b}\rightarrow \Omega_{c}D^{-}$&$-0.333\pm0.092$&$0.771\pm0.226$&$-0.047\pm0.015$&$-0.021\pm0.006$&$0.468\pm0.120$&$0.517\pm0.150$\\
      			$\Omega_{b}\rightarrow \Omega_{c}D_{s}^{-}$
      			&$-0.339\pm0.093$ &$0.782\pm0.230$& $-0.048\pm0.015$& $-0.021\pm0.007$&$0.475\pm0.120$&$0.525\pm0.151$\\
      			$\Omega_{b}\rightarrow \Omega_{c}\rho^{-}$&$-0.297\pm0.082$&$0.694\pm0.204$&$-0.039\pm0.012$&$-0.020\pm0.007$&$0.419\pm0.112$&$0.460\pm0.122$\\
      			$\Omega_{b}\rightarrow \Omega_{c}K^{*-}$& $-0.299\pm0.082$&$0.699\pm0.206$&$-0.040\pm0.013$&$-0.020\pm0.006$&$0.422\pm0.103$&$0.464\pm0.124$\\
      			$\Omega_{b}\rightarrow \Omega_{c}D^{*-}$& 
      			$-0.341\pm0.094$& $0.786\pm0.231$&$-0.049\pm0.015$&$-0.021\pm0.007$&$0.478\pm0.119$&$0.528\pm0.141$\\
      			$\Omega_{b}\rightarrow \Omega_{c}D_{s}^{*-}$ 
      			&$-0.347\pm0.096$ &$0.799\pm0.235$& $-0.050\pm0.016$&$-0.021\pm0.006$&$0.487\pm0.130$&$0.538\pm0.144$
      		\end{tabular}
      	\end{ruledtabular}
      	\label{formfactornon}
      \end{table}
By substituting the form factors into the decay rate expressions given in \ref{decayformula}, the decay rates and branching ratios for the pseudoscalar($\pi, K,D, D_s$) and vector ($\rho,K^*, D^*, D_s^*$) mesons are obtained and reported  in Tables \ref {DECAY} and \ref{BR}.  These tables include the other predictions in the literature, as well.
 The obtained decay rates and branching ratios in the present study,  considering the uncertainties, are overall  in good agreement with the other existing predictions. 
It is worth mentioning that different references use different form factors calculated via different approaches and consider different topological contributions:  The differences among the presented results for some channels  can be attributed in the first stage to the different form factors used in these references as well as different topologies  considered.  The choice of form factors  and their behavior in terms of $ q^2 $ as well as their values at $ q^2=m^2_{meson} $ play important role in enhancing our results for the  branching ratios in some channels in Table \ref{BR}. 
References  \cite{Li:2021kfb,Zhao:2018zcb} and \cite{Chua:2019yqh,Chua:2019lgb} calculate the related form factors using the light-front method and examine nonleptonic decays only at the tree level (color allowed). References \cite{Gutsche:2018utw,Ivanov:1997hi} employ the covariant confined quark model and investigate the decays at the tree level (color allowed), incorporating QCD penguin operators. Reference \cite{Cheng:1996cs} uses the non-relativistic quark model to determine form factors and calculates the decay rates at the tree level including color-suppressed contributions. The
Reference \cite{Patel:2025gbw} employs the independent quark model within a relativistic framework and examines the corresponding decays at the tree level (color allowed).
In contrast,
the present work provides the most complete analysis of the nonleptonic decay $\Omega_b \to \Omega_c$ incorporating tree-level,
color-suppressed, and penguin topologies. To illustrate the impact of each topology on the decay rates, the partial
decay rates corresponding to individual topological contributions for each decay channel are listed in Table  \ref{DECAYtopo}.
This decomposition allows for a clear assessment of the relative importance of different topologies across various
decay modes. As shown, although the pure decay rates from color-suppressed and penguin diagrams are small,
they can significantly affect the total decay rates and should not be neglected. These results provide a comprehensive reference for analyzing asymmetry effects and their role in different decay
channels from the viewpoint of CP violation effects.
\begin{table}[bth]
\caption{Decay widths (in $\times10^{15}\mathrm{GeV}$) for the nonleptonic $\Omega_b\to\Omega_c M$ transition at different channels.}\label{DECAY}
\begin{ruledtabular}
\begin{tabular}{|c|c|c|c|c|}
             &This work & \cite{Ivanov:1997hi}& \cite{Zhao:2018zcb} &\cite{Cheng:1996cs}    \\
\hline
$\Gamma~[\Omega_b\to\Omega_c \pi]$ &$1.809\pm0.925$& $2.32$  & $1.68$ & $2.09$\\
\hline
$\Gamma~[\Omega_b\to\Omega_c K]$ & $0.171\pm 0.085$&--& $0.136$ &- \\
\hline
$\Gamma~[\Omega_b\to\Omega_c D]$ & $0.202\pm0.099$ &-& $0.226$ &-\\
\hline
$\Gamma~[\Omega_b\to\Omega_c D_s]$ &$6.185\pm3.045$&-& $7.17$ & $7.62$\\
\hline
$\Gamma~[\Omega_b\to\Omega_c \rho]$ &$5.047\pm2.495$ &-& $4.54$ & $2.73$ \\
\hline
$\Gamma~[\Omega_b\to\Omega_c K^*]$ &$0.302\pm0.149$&--&$0.228$ &-\\ 
\hline
$\Gamma~[\Omega_b\to\Omega_c D^*]$ & $0.274\pm0.133$&-& $0.214$ &-\\
\hline
$\Gamma~[\Omega_b\to\Omega_c D_s^*]$ & $7.446\pm3.626$ &-& $4.90$ &$2.45$\\
\hline
\end{tabular}
\end{ruledtabular}
\end{table}
\begin{table}[bth]
\caption{Branching ratios  in $\times10^{-3}$ for the  nonleptonic $\Omega_b\to\Omega_c M$ transition at different channels.}\label{BR}
\begin{ruledtabular}
\begin{tabular}{|c|c|c|c|c|c|c|c|c|}
             &This work &\cite{Gutsche:2018utw}&\cite{Li:2021kfb}&  \cite{Zhao:2018zcb}& \cite{Cheng:1996cs} &\cite{Patel:2025gbw} &\cite{Chua:2019yqh} &\cite{Chua:2019lgb}   \\
\hline
$Br~[\Omega_b\to\Omega_c \pi]$ &$4.657\pm2.305$ &$1.88$ & $2.82$ &$4.00$& $4.92$& $2.54$ &  $1.33^{+0.62}_{-0.46}$&$1.10^{+0.85}_{-0.55}$\\
\hline
$Br~[\Omega_b\to\Omega_c K]$ &$0.427\pm0.211$ &- & $0.22$&$0.326$ & -&$0.20$ &$0.10^{+0.05}_{-0.03}$&$0.08^{+0.07}_{-0.04}$\\
\hline
$Br~[\Omega_b\to\Omega_c D]$ &$0.503\pm0.247$&$0.24$ & $0.52$ & $0.636$ &- & $0.46$ & $0.18^{+0.11}_{-0.08}$&$0.15^{+0.14}_{-0.08}$\\
\hline
$Br~[\Omega_b\to\Omega_c D_s]$ &$15.411\pm7.589$&-& $13.5$&$17.1$ & $17.9$ & $12.2$ & $4.75^{+2.85}_{-1.97}$&$4.03^{+3.72}_{-2.21}$ \\
\hline
$Br~[\Omega_b\to\Omega_c \rho]$ &$12.575\pm6.216$ &$5.43$& $7.92$&$10.8$  & $12.8$ & $7.43$ & $1.63^{+1.17}_{-0.75}$ &$3.07^{+2.41}_{-1.53}$\\
\hline
$Br~[\Omega_b\to\Omega_c K^*]$ &$0.751\pm0.380$&- & $0.41$  & $0.544$ &-& $0.32$ & $0.082^{+0.058}_{-0.037}$&$0.16^{+0.12}_{-0.08}$\\ 
\hline
$Br~[\Omega_b\to\Omega_c D^*]$ & $0.684\pm0.334$&$0.30$& $0.48$& $0.511$ & - & $0.39$ & $0.088^{+0.055}_{-0.035}$&$0.16^{+0.13}_{-0.08}$ \\
\hline
$Br~[\Omega_b\to\Omega_c D_s^*]$ &$18.553\pm9.036$&-& $9.73$& $11.7$  & $9.73$& $9.44$ & $1.84^{+1.11}_{-0.71}$ &$3.18^{+2.69}_{-1.61}$ \\
\hline
\end{tabular}
\end{ruledtabular}
\end{table}
\begin{table}[bth]
\caption{The pure contribution of the decay rate corresponding to each topology (in $\times10^{15}\mathrm{GeV}$) for $\Omega_b\to\Omega_c M$ transition at different channels.}\label{DECAYtopo}
\begin{ruledtabular}
\begin{tabular}{|c|c|c|c|c|}
             &Tree level   &Color-suppressed& Penguin & Total decay    \\
\hline
$\Gamma~[\Omega_b\to\Omega_c \pi]$ & $1.869$ & -  & - & $1.869$\\
\hline
$\Gamma~[\Omega_b\to\Omega_c \rho]$ &$5.047$ &-& - & $5.047$\\
\hline
$\Gamma~[\Omega_b\to\Omega_c K]$ &$0.139$& $0.002$ & -&$0.171$  \\
\hline
$\Gamma~[\Omega_b\to\Omega_c K^*]$ &$0.244$ &$0.003$ &- &$0.302$\\ 
\hline
$\Gamma~[\Omega_b\to\Omega_c D]$ & $0.158$ &-& $0.003$ & $0.202$\\
\hline
$\Gamma~[\Omega_b\to\Omega_c D^*]$ & $0.255$ &-& $0.0004$ &$0.274$\\
\hline
$\Gamma~[\Omega_b\to\Omega_c D_s]$ &$4.744$& $0.059$ & $0.004$ & $6.185$\\
\hline
$\Gamma~[\Omega_b\to\Omega_c D_s^*]$ & $5.933$ & $0.074$ & $0.0004$ & $7.446$\\
\hline
\end{tabular}
\end{ruledtabular}
\end{table}
\section{Conclusion}~\label{Sec5}
In this study,  we investigated the non-leptonic two-body decays of $\Omega_b \to \Omega_c$ in both the pseudoscalar and vector channels including eight allowed mesonic modes.  We examined all possible contributions to the related amplitudes incorporating the  tree-level,
color-suppressed, and penguin topologies for these decays and computed the decay amplitudes for all these channels.  The decay rate and branching ratio  for  each decay were calculated considering all possible contributions from the tree level,  color-suppressed and penguin configurations, and the results were compared with predictions from other existing theories. The contributions of each topology for different mesonic channels were also given separately:  Though the contributions from the color-suppressed and penguin diagrams are small, they have  significant  effects to the total decay rates and branching fractions.  Our results can be used in the ongoing and future experiments: The results can play significant role in exact determination of the lifetime as well as other properties and internal structure of the relatively  less-known $\Omega_b$ baryon as one of the important members of the single heavy spin-1/2 baryon family.

\section*{ACKNOWLEDGMENTS} 
K. Azizi thanks Iran national science foundation (INSF) for the partial financial support supplied under the elites Grant No. 40405095. 
\appendix \label{wilson}
\section{The penguin-operator coefficients}
The QCD penguin operator $O_3$ is written as follows:
\begin{eqnarray}
&&O_{3}=(\bar{d_{i}}b_{i})_{V-A}\sum_{q}(\bar{q}_{j}q_{j})_{V-A}.  \\ \notag
\end{eqnarray}
This representation of the $O_3$ operator is unsuitable and cannot produce $\Omega_b \to \Omega_c D$. 
Fierz transformations rearrange bilinears in four-fermion interactions, expressed as:
\begin{eqnarray}
(\bar{\psi}_1 \Gamma^A \psi_2)(\bar{\psi}_3 \Gamma_A \psi_4) = \sum_B C^{AB} (\bar{\psi}_1 \Gamma^B \psi_4)(\bar{\psi}_3 \Gamma_B \psi_2),
\end{eqnarray}
where $\Gamma^A$ form a basis of Dirac matrices and $C^{AB}$ are Fierz coefficients.
The coefficients  are derived from
\begin{eqnarray}
 C^{AB} =\frac{1}{16 N_{B}}  \mathrm{Tr}(\Gamma^A_a \Gamma^B_b \Gamma^A_a \Gamma^B_b),
 \end{eqnarray}
  where $ N_{B} $ is the number of gamma matrices in each group $ B $:  In our case  $ N_{B} =4$ .
After a Fierz transformation, it takes the form below:
\begin{eqnarray}
&&O_{3}=(\bar{d_{i}}c_{j})_{V-A}(\bar{c}_{j}b_{i})_{V-A}.\\ \notag
\end{eqnarray}
A further transformation then places the colors $i$ and $j$ adjacent to each other:
\begin{eqnarray}
&&O_{3}=(\bar{d_{i}}c_{i})_{V-A}(\bar{c}_{j}b_{j})_{V-A}, \\ \notag
\end{eqnarray}
where the color rearrangement introduces a factor of 1/3, and the operator $O_3$ ultimately takes the form:
\begin{eqnarray}
&&O_{3}=\frac{1}{3} C_3 O_{1} .
\end{eqnarray}
For the operator $O_4$, we apply the Fierz transformation once to place the quarks adjacent to each other, yielding:
\begin{eqnarray}
&&O_{4}=(\bar{d_{i}}b_{j})_{V-A}\sum_{q}(\bar{q}_{j}q_{i})_{V-A},  \\ \notag
&&O_{4}=(\bar{d_{i}}c_{i})_{V-A}(\bar{c}_{j}b_{j})_{V-A}, \\ \notag
&&O_{4}= C_4 O_{1} .
\end{eqnarray}
The operators $O_5$ to $O_8$ are of the form $(V-A)\otimes (V+A)$,  which the following transformations are employed. The operator $O_6$, for instance,   is initially given by:
\begin{eqnarray}
&&O_{6}=(\bar{d_{i}}b_{j})_{V-A}\sum_{q}(\bar{q}_{j}q_{i})_{V+A},  \\ \notag
\end{eqnarray}
we have:
\begin{eqnarray}
&&\gamma_\mu (1-\gamma_5) \otimes \gamma_\mu(1+\gamma_5) \to -2 (1-\gamma_5)(1+\gamma_5) .\\ \notag
\end{eqnarray}
After applying the Fierz transformation and factorization, the matrix element factorizes into:
\begin{eqnarray}
\left\langle \Omega_{c}M\left|O_6\right|\Omega_{b}\right\rangle =-2\sum_{q'}\left\langle M\left|\bar q(1+\gamma_5) q'\right|0\right\rangle  \left\langle \Omega_{c}\left|\bar c (1-\gamma_5) b\right|\Omega_{b}\right\rangle .
\end{eqnarray}
We can begin with the meson matrix element:
\begin{eqnarray}
&&\langle  D |\bar d \gamma_\mu(1-\gamma_5) c |0 \rangle=if_D q_\mu,\\ \notag
\end{eqnarray}
at which we contract both sides with $ q^\mu$, where $q^\mu$ is the meson four-momentum. Then, $q^2$ is replaced by $ m_{M}^2$, the squared mass of the meson, since the on-shell condition for the meson gives $q^2 = m_M^2$:
\begin{eqnarray}
&&\langle  D |\bar d ~ \slashed{q} (1-\gamma_5) c |0 \rangle= i f_D m^2_D.\\ \notag
\end{eqnarray}
We have $q_D = q_d + q_c$, where $q_D$ is the four-momentum of the $ D$ meson and $ q_d$  and $q_c$ are the four-momenta of the constituent $d$ and $c$ quarks, respectively:
\begin{eqnarray}
&&\langle  D |\bar d ~( \slashed{p}_q+\slashed {p}_{q'}) (1-\gamma_5) c |0 \rangle= i f_D m^2_D.\\ \notag
\end{eqnarray}
By substituting from the Dirac equation:
\begin{eqnarray}
&&\langle  D |\bar d  (1+\gamma_5) c |0 \rangle= -i f_D \frac{ m^2_D }{(m_d+m_{c})}.
\end{eqnarray}
The same argument is applied to the baryon matrix element.
\section{The  squared amplitudes}\label{squared amplitudes}
The squared amplitude for the decay $\Omega_b \to \Omega_c P$ is given by:
\begin{eqnarray}
&&|\mathcal{A}|^2=(\frac{G_F}{\sqrt2} f_P C_{\rm eff})^2\frac{1}{2 m^2_{\Omega_b} m^2_{\Omega_c}} \Bigg\{-F_3^2 m^2_{\Omega_b}m_z (m^2_{\Omega_b} + 3 m^2_{\Omega_c} -
       q^2)^2 +  G_3^2 m^2_{\Omega_b}m_{z'} (m^2_{\Omega_b} + 3 m^2_{\Omega_c} - q^2)^2 + \notag\\
  && 2 G_3 m_{\Omega_b} m_{\Omega_c} \Big[2 G_1 m_{\Omega_b} m_+ + 
      G_2 (-3 m^2_{\Omega_b}  - m')\Big] 
      \Big[m^4_{\Omega_b}  +  2 m^3_{\Omega_b}  m_{\Omega_c}  + 3 m^4_{\Omega_c}  + \notag\\
    && m^2_{\Omega_b}  (4 m^2_{\Omega_c}  - 2 q^2) - 4 m^2_{\Omega_c}  q^2 + q^4 + 
      m_{\Omega_b}  (6 m^3_{\Omega_c} - 2 m_{\Omega_c} q^2)\Big] + 
   2 F_3 m_{\Omega_b} m_{\Omega_c}\Big [2 F_1 m_{\Omega_b} m_- + \notag\\
    &&  F_2 (3 m^2_{\Omega_b} + m')\Big] \Big[m^4_{\Omega_b} - 
      2 m^3_{\Omega_b} m_{\Omega_c}+ 3 m^4_{\Omega_c} + 
      m^2_{\Omega_b} (4m^2_{\Omega_c} - 2 q^2) - 4 m^2_{\Omega_c} q^2 + q^4 + \notag\\
     && m_{\Omega_b} (-6 m^3_{\Omega_c} + 2 m_{\Omega_c} q^2)\Big] + 
m^2_{\Omega_c}  \Bigg[-4 F_1^2 m^2_{\Omega_b} m_-^2 m_z - 
      F_2^2 (3 m^2_{\Omega_b}  + m')^2 
      m_z + m_{z'} \notag\\
      &&\bigg(2 G_1 m_{\Omega_b}  m_+ + 
        G_2 (-3 m^2_{\Omega_b}  - m')\bigg)^2 - 
      4 F_1 F_2 m_{\Omega_b} m_- \Big(3 m^4_{\Omega_b}  - 
         6 m^3_{\Omega_c}  m_{\Omega_c}  + \notag\\
       &&  4  m^2_{\Omega_b} m'+ m'^2 - 
         2 m_{\Omega_b} m_{\Omega_c}m'\Big)\Bigg]\Bigg\}.
         \end{eqnarray}
      For the decay $\Omega_b \to \Omega_c V$, we have:
      \begin{eqnarray}
&&|\mathcal{A}|^2=(\frac{G_F}{\sqrt2} f_V m_V C_{\rm eff})^2
     \frac{ 1}{2  m^2_{\Omega_b}  m^2_{\Omega_c} q^2}\Bigg\{2 G_3 m_{\Omega_b}  m_{\Omega_c}
      m_{z'}
     \Bigg[-G_2\bigg (3 m^4_{\Omega_b}  + 2  m^2_{\Omega_b}(5 m^2_{\Omega_c} - 3 q^2) + 
        3 m'^2\bigg) + \notag\\
  &&   2 G_1 m_{\Omega_b} \bigg (m^3_{\Omega_b}  +  m^2_{\Omega_b} m_{\Omega_c} + 
        3 m_{\Omega_b}   m^2_{\Omega_c}+ 3 m^3_{\Omega_c}  - m_{\Omega_b}  q^2 - 
        3 m_{\Omega_c}  q^2\bigg)\Bigg] + \notag\\
 && 2 F_3 m_{\Omega_b}  m_{\Omega_c}  m_z
   \Bigg[F_2 \bigg(3 m^4_{\Omega_b} + 2  m^2_{\Omega_b} (5 m^2_{\Omega_c} - 3 q^2) + 
        3 m'^2\bigg) + \notag\\
   &&  2 F_1m_{\Omega_b}  (m^3_{\Omega_b}  - m^2_{\Omega_b} m_{\Omega_c}  + 
        3 m_{\Omega_b}   m^2_{\Omega_c} - 3 m^3_{\Omega_c}  - m_{\Omega_b}  q^2 + 
        3 m_{\Omega_c}  q^2)\Bigg] + \notag\\
 && F_3^2 m^2_{\Omega_b} \Bigg[-m^6_{\Omega_b}  + 
     2 m^5_{\Omega_b}  m_{\Omega_c}  - m'^2 (9  m^2_{\Omega_c} - q^2) +
      m^4_{\Omega_b}  (-7 m^2_{\Omega_c}  + 3 q^2) + 
     4 m^3_{\Omega_b}  (3 m^3_{\Omega_c}  - m_{\Omega_c}  q^2) + \notag\\
    && 2m_{\Omega_b} m_{\Omega_c}  (9m^4_{\Omega_c}  - 10 m^2_{\Omega_c}  q^2 + q^4) - 
     3 m^2_{\Omega_b}  (5 m^4_{\Omega_c}  - 6  m^2_{\Omega_c} q^2 + q^4)\Bigg] + \notag\\
&&  G_3^2 m^2_{\Omega_b}  \Bigg[m^6_{\Omega_b} + 2 m^5_{\Omega_b}  m_{\Omega_c}  + 
   m^4_{\Omega_b}  (7 m^2_{\Omega_c} - 3 q^2) + m'^2 (9  m^2_{\Omega_c} - q^2) + 
     4 m^3_{\Omega_b} (3 m^3_{\Omega_c} - m_{\Omega_c} q^2) + \notag\\
    && 2 m_{\Omega_b} m_{\Omega_c} (9 m^4_{\Omega_c} - 10 m^2_{\Omega_c} q^2 + q^4) + 
     3 m^2_{\Omega_b} (5 m^4_{\Omega_c} - 6 m^2_{\Omega_c} q^2 + q^4)\Bigg] + \notag\\
 && m^2_{\Omega_c} \Bigg[4 G_1^2 m^6_{\Omega_b}- 12 G_1 G_2 m^6_{\Omega_b} + 
     9 G_2^2 m^6_{\Omega_b}+ 16 G_1^2 m^5_{\Omega_b} m_{\Omega_c} - 
     36 G_1 G_2 m^5_{\Omega_b} m_{\Omega_c} + 18 G_2^2 m^5_{\Omega_b} m_{\Omega_c} + \notag\\
     &&24 G_1^2 m^4_{\Omega_b} m^2_{\Omega_c} - 40 G_1 G_2 m^4_{\Omega_b} m^2_{\Omega_c} + 
     15 G_2^2 m^4_{\Omega_b}m^2_{\Omega_c} + 16 G_1^2 m^3_{\Omega_b} m^3_{\Omega_c}- 
     24 G_1 G_2 m^3_{\Omega_b} m^3_{\Omega_c} + 12 G_2^2 m^3_{\Omega_b}m^3_{\Omega_c} + \notag\\
    && 4 G_1^2 m^2_{\Omega_b} m^4_{\Omega_c} - 12 G_1 G_2 m^2_{\Omega_b} m^4_{\Omega_c} + 
     7 G_2^2 m^2_{\Omega_b} m^4_{\Omega_c} - 4 G_1 G_2 m_{\Omega_b} m^5_{\Omega_c} + 
     2 G_2^2 m_{\Omega_b} m^5_{\Omega_c} + G_2^2 m^6_{\Omega_c} + \notag\\
    && 4 G_1^2 m^4_{\Omega_b} q^2 + 24 G_1 G_2 m^4_{\Omega_b} q^2- 
    19 G_2^2 m^4_{\Omega_b} q^2 - 40 G_1^2 m^3_{\Omega_b} m_{\Omega_c} q^2+ 
     40 G_1 G_2 m^3_{\Omega_b}m_{\Omega_c} q^2 - 
     20 G_2^2 m^3_{\Omega_b} m_{\Omega_c}q^2+ \notag\\
    && 4 G_1^2 m^2_{\Omega_b} m^2_{\Omega_c} q^2 + 
     24 G_1 G_2 m^2_{\Omega_b} m^2_{\Omega_c} q^2 - 
     18 G_2^2 m^2_{\Omega_b} m^2_{\Omega_c} q^2 + 
     8 G_1 G_2 m_{\Omega_b} m^3_{\Omega_c} q^2 - 4 G_2^2 m_{\Omega_b} m^3_{\Omega_c} q^2- 
     3 G_2^2 m^4_{\Omega_c} q^2\notag\\ 
     &&- 8 G_1^2 m^2_{\Omega_b} q^4 - 
     12 G_1 G_2 m^2_{\Omega_b} q^4 + 11 G_2^2 m^2_{\Omega_b} q^4 - 
     4 G_1 G_2 m_{\Omega_b}  m_{\Omega_c}  q^4 + 2 G_2^2 m_{\Omega_b}  m_{\Omega_c} q^4 + 
     3 G_2^2 m^2_{\Omega_c} q^4 - G_2^2 q^6 - \notag\\
    && 4 F_1 F_2 m_{\Omega_b}  \bigg(3 m^5_{\Omega_b}- 9 m^4_{\Omega_b} m_{\Omega_c}+ 
        2 m^3_{\Omega_b} (5 m^2_{\Omega_c} - 3 q^2) + 
        3 m_{\Omega_b}  m'^2 - 
        m_{\Omega_c} m'^2 + \notag\\
    &&    m^2_{\Omega_b} (-6 m^3_{\Omega_c} + 10 m_{\Omega_c} q^2)\bigg) - 
     4 F_1^2 m^2_{\Omega_b}\bigg (m^4_{\Omega_b} - 4 m^3_{\Omega_b} m_{\Omega_c} + m^4_{\Omega_c} + 
        m^2_{\Omega_c} q^2 - 2 q^4 + m^2_{\Omega_b} (6 m^2_{\Omega_c} + q^2) + \notag\\
      &&  m_{\Omega_b} (-4 m^3_{\Omega_c} + 10 m_{\Omega_c} q^2)\bigg) + 
     F_2^2\bigg (-9 m^6_{\Omega_b} + 18 m^5_{\Omega_b} m_{\Omega_c}  + 
        2 m_{\Omega_b}  m_{\Omega_c}  (m^2_{\Omega_c}  - q^2)^2 - (m^2_{\Omega_c}  - 
          q^2)^3+\notag\\
          &&  m^4_{\Omega_b}  (-15 m^2_{\Omega_c}  + 19 q^2) + 
        4 m^3_{\Omega_b}  (3m^3_{\Omega_c}  - 5 m_{\Omega_c}  q^2) + 
        m^2_{\Omega_b}  (-7 m^4_{\Omega_c}  + 18 m^2_{\Omega_c}  q^2 - 11 q^4)\bigg)\Bigg]\Bigg\},
         \end{eqnarray}
           where
         \\ 
        $m_z=m^2_{\Omega_b} -  2 m_{\Omega_b} m_{\Omega_c} + m^2_{\Omega_c} - q^2$,
      \\
      $m_{z'}=m^2_{\Omega_b} + 2 m_{\Omega_b} m_{\Omega_c} + m^2_{\Omega_c} -  q^2$,
      \\
      $m_+=m_{\Omega_b} + m_{\Omega_c}$,
      \\
      $m_-=m_{\Omega_b} - m_{\Omega_c}$,
      \\
      $m'=m^2_{\Omega_c}-q^2$.
      \\
\label{sec:Num}

\end{document}